\documentclass[a4paper,11pt]{article}
\usepackage{pos}

\title{Ultra-high-energy neutrino detection with radio antennas in the ground-based observatory}
\author*[a]{Baobiao Yue}
\author[a]{Karl-Heinz Kampert}
\author[a]{Julian Rautenberg}

\affiliation[a]{Bergische Universität Wuppertal,\\
  Gaußstraße 20, Wuppertal, Germany}

\emailAdd{bayue@uni-wuppertal.de}
\emailAdd{kampert@uni-wuppertal.de}
\emailAdd{julian.rautenberg@uni-wuppertal.de}

\abstract{
Ultra-high-energy (UHE) neutrinos are unique cosmic messengers that can traverse cosmological distances unattenuated, offering direct insight into the most energetic processes in the universe. 
Radio detection promises significant advantages for detecting highly inclined air showers induced by UHE neutrinos, including a larger exposure range compared to particle detectors, which is due to minimal atmospheric attenuation of radio signals combined with good reconstruction precision. Furthermore, this technique improves the air shower longitudinal reconstruction, which can be used to identify neutrinos with their first interaction far below the top of the atmosphere.
In this work, we present a method for identifying UHE neutrinos using radio antennas deployed in ground-based observatories. We introduce a reconstruction algorithm based on the radio emission maximum ($X^{\text{radio}}_{\text{max}}$) and demonstrate its power in distinguishing deeply developing neutrino-induced showers from background cosmic rays. Using the Pierre Auger Observatory as a case study, we use the simulations of $\nu_e$-CC-induced air showers and evaluate the trigger efficiency, reconstruction performance, and resulting effective area.
Our results show that radio detection significantly enhances the sensitivity to very inclined showers above 1~EeV, complementing traditional surface detectors. This technique is highly scalable and applicable to future radio observatories such as GRAND. The proposed reconstruction and identification strategy provides a pathway toward achieving the sensitivity needed to detect UHE neutrinos.
}
\ConferenceLogo{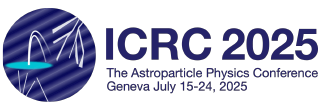}

\FullConference{39th International Cosmic Ray Conference (ICRC2025)\\
 15–24 July 2025\\
Geneva, Switzerland\\}
\begin{document}
\maketitle

%\section{Introduction}
%Ultra-high-energy (UHE) neutrinos offer crucial insights into cosmic accelerators. As messengers that travel undisturbed across cosmic distances, their detection provides a complementary channel to cosmic ray and gamma-ray observations. Detecting UHE neutrinos with radio antennas leverages the long-range propagation of radio waves, minimal attenuation in the atmosphere, and broad coverage of ground-based observatories.

%The Pierre Auger Observatory, with its radio antenna extensions, presents a promising platform for this purpose. In particular, inclined air showers generated by down-going neutrinos are well-suited for radio detection due to their extended development in the atmosphere.

\section{Introduction}

Ultra-high-energy (UHE) neutrinos, with energies exceeding $10^{17}$~eV, are important messengers for uncovering the origins of the most energetic cosmic rays in the universe. Due to their weak interactions and lack of electric charge, neutrinos propagate over cosmological distances without deflection or significant energy loss, carrying the information of the direction and the radiant energy from their sources. 
% They can be produced in astrophysical accelerators, such as AGN, GRBs, 
% such as active galactic nuclei (AGN), gamma-ray bursts (GRBs), 
% or via interactions of UHE cosmic rays (UHECR) with cosmic background radiation, giving rise to the so-called cosmogenic or GZK neutrinos.

A growing global effort is underway to detect these elusive particles. The IceCube Neutrino Observatory has pioneered neutrino astronomy by observing a diffuse astrophysical neutrino flux in the TeV–PeV range.
% However, its stringent upper limits at EeV energies~\cite{IceCubeDiffuseLimit} indicate a steeply falling flux, motivating the search for complementary detection techniques. 
Most recently, the KM3NeT reported the detection of a $\sim$220~PeV neutrino candidate~\cite{KM3NeTEvent}, the most energetic such event to date. 
Its origin remains unknown, which has aroused more interest in UHE neutrino detection.
% The source of it is still unknown, which motivates 
% Its energy and near-horizontal trajectory suggest it may originate from a distinct population of accelerators, or perhaps represent the first observed cosmogenic neutrino.

The detection of UHE neutrinos requires enormous effective volumes and advanced background rejection. 
A promising avenue involves large-area radio antenna arrays deployed on the ground or embedded in natural media. 
The ground based radio detection exploits the coherent radio emission generated by the geomagnetic deflection of charged particles, or the charge excess (Askaryan effect) in an extensive air shower. 
Radio signals are minimally attenuated in the atmosphere and offer nanosecond timing information, enabling precise reconstruction of the shower geometry and energy. 

% Ground-based observatories, such as the Pierre Auger Observatory, illustrate this approach. 
% % Though it was originally designed for cosmic-ray detection, 
% The Pierre Auger Observatory has conducted leading UHE neutrino searches using its surface detector array to detect highly inclined showers. 
% The Earth-skimming and down-going channels have yielded most stringent limits on diffuse UHE neutrino fluxes~\cite{AugerNeutrinoSearch2024}. 
% Building on this foundation, radio detection offers several advantages including increased exposure to inclined showers, improved reconstruction of the shower maximum ($X_{\mathrm{max}}$), and sensitivity to neutrino interaction deep within the atmosphere. 
% Additional works on this topic can be found e.g.\ in \cite{Washington}.

In this study, we examine the potential of radio detection to improve UHE neutrino searches at ground-based observatories. Using the Pierre Auger Observatory as a case study, we evaluate the performance of radio-based triggering and reconstruction for inclined showers. We also describe methods to identify neutrino-induced air showers via $X_{\mathrm{max}}^{\mathrm{radio}}$ and estimate the sensitivity achievable with a standalone radio array. Our findings show how radio detection can complement existing techniques and expand the capabilities of neutrino observatories to the EeV regime. Additional works on this topic can be found e.g.\ in \cite{Washington}.

\section{Radio detection for neutrino-induced air showers}
Downward-going neutrinos can interact within the atmosphere, producing extensive air showers (EAS). These interactions can occur at slant depths much greater than those of cosmic rays. The key parameter that distinguishes neutrinos from hadronic cosmic rays is the first interaction point, $X_1$, or the position of maximum radio emission, $X^{\text{radio}}_{\text{max}}$. 
Because radio radiation is coherent, we can reconstruct the position of maximum radio emission, $X^{\text{radio}}_{\text{max}}$, using the timing of the maximum pulse in the radio antennas. This enables us to distinguish neutrinos with deep atmospheric interactions from inclined UHECRs.

Earth-skimming tau neutrinos can create tau-decay-induced air showers close to the ground coming from directions a few degrees below the horizon. If the air shower is close to the observatory array, the particle detector can recognize it as a young shower due to its obvious electromagnetic signal. 
Using radio interferometry \cite{LOPES}, the shower axis can be reconstructed with high precision. This allows for the identification of upward-going neutrinos traversing the Earth, as well as downward-going tau neutrinos emerging from interactions with mountains.
%With the radio interferometry technique \cite{RadidoInterferometry}, the shower axis could be reconstructed precisely, which could be used to identify the   upward-going neutrino through the Earth, which could also be used as well to identify the downward-going tau neutrinos through the mountain. 

In this work, we will focus on the detection of down-going $\nu_e$-CC interactions using radio antennas in ground-based radio arrays to demonstrate the potential of radio antennas for neutrino detection.  
%The ground based array should be able to detect 

%Radio emission arises primarily from geomagnetic and Askaryan effects. The coherent radio signals, especially in the 30–80~MHz band, are minimally affected by atmospheric absorption, offering high-fidelity shower reconstruction.

\section{Simulations and neutrino trigger efficiency}
\begin{figure}
    \centering
    \includegraphics[width=0.45\linewidth]{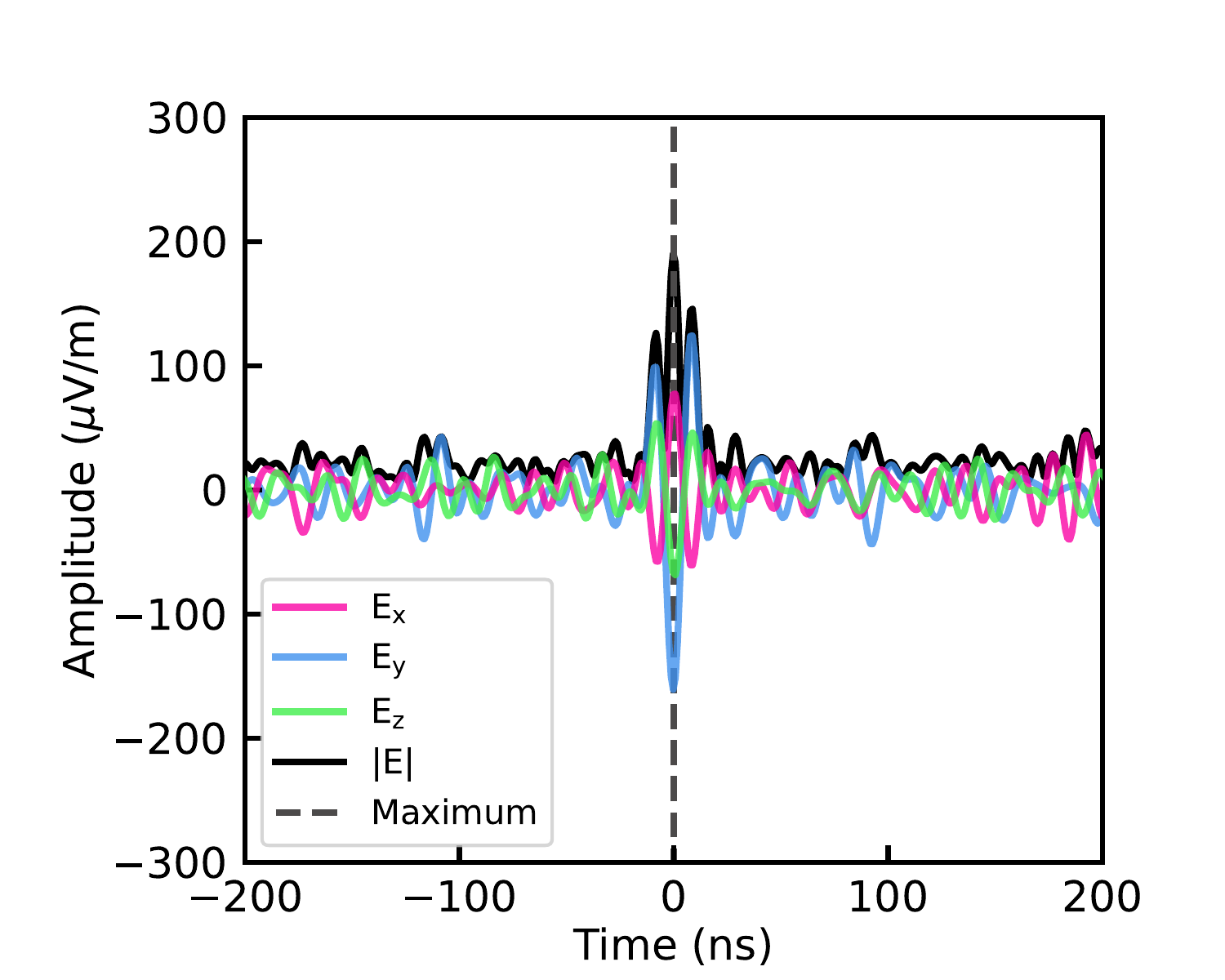}
    \includegraphics[width=0.45\linewidth]{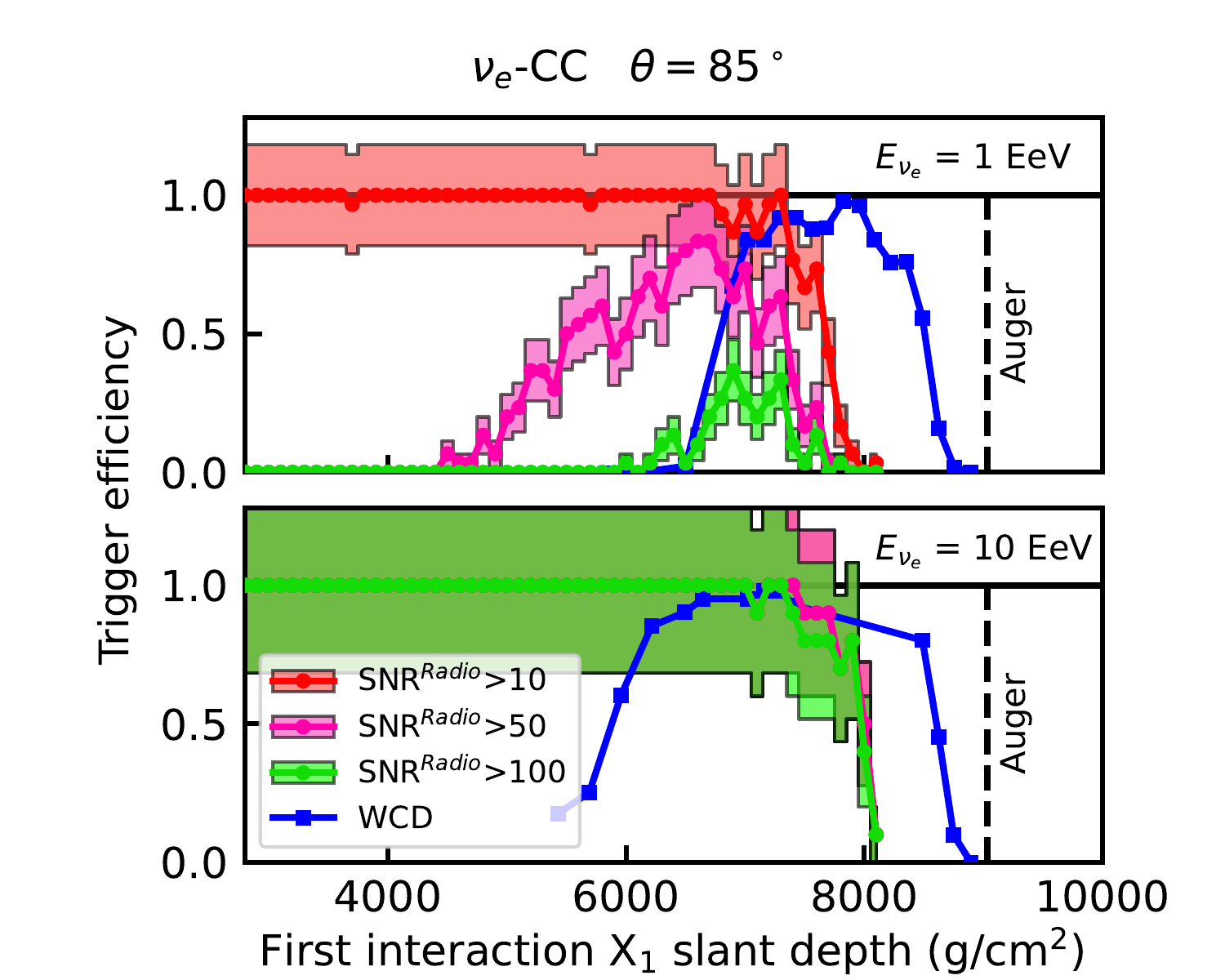}
    \caption{Left: electric field simulation with 30-80 MHz bandpass filter and white noise. Right: $\nu_e$-CC trigger efficiency for radio antennas at different thresholds compared to water Cherenkov detectors (WCD). $X_1=0$ represents the top of the atmosphere.}
    \label{fig:RadioSignal}
\end{figure}

Simulations of down-going neutrino $\nu_e$ charged current (CC) interactions from CoREAS \cite{huege2013} were used to study the detection sensitivity. 
These simulations contain the time-dependent the electric field vector from the air showers at specific antenna positions. 
The CoREAS simulations use 1.5 km spacing to match the configuration of the Pierre Auger Observatory. 
To mimic the frequency response of the radio antenna and the analog filter-amplifier chain, the simulations use a 30-80\,MHz bandpass filter from common radio detectors like the Pierre Auger Observatory.
We generate an uncertainty of $\sigma^\text{GPS}_t=5$\,ns for the local time of the GPS receiver in each individual antenna without external synchronization signals.
Additionally, we simulate a unpolarized white noise with the root mean square (RMS) of about 25\,\textmu V/m, denoted as RMS$_\mathrm{noise}$, in the radio signal traces to mimic the realistic situation.
% the radio background from the Milky Way \cite{RadidoInterferometry}, especially in the frequency range between 30–80\,MHz.
The left side of Fig.\,\ref{fig:RadioSignal} presents the electric field strengths of the $x$, $y$, and $z$ polarizations with a 30-80\,MHz bandpass filter and the white noise.
%( ($\Theta$, $\Phi$, $r$) is the direction vector of the incident electric field in the spherical coordinate system with the antenna as the origin). % perpendicular to the radio wave propagation. 
We define a trigger threshold value, denoted as T$_\mathrm{trigger}^\mathrm{SNR}$, for the signal-to-noise ratio (SNR), written as follows:
\begin{equation}
    \mathrm{SNR} = \left|\frac{\mathrm{E}}{\mathrm{RMS}_\mathrm{noise}}\right|^2\,.
\end{equation}
% where $\sqrt{3}$ represents the summation of the white noise summation from three perpendicular polarizations to mimic the radio background from the Milky Way, especially in the frequency range between 30–80\,MHz.
If the SNR is greater than T$_\mathrm{trigger}^\mathrm{SNR}$, the antenna is tagged as triggered.
We define three cases: an ideal case with T$_\mathrm{trigger}^\mathrm{SNR}=10$, an intermediate case with T$_\mathrm{trigger}^\mathrm{SNR}=50$, and a conservative case with T$_\mathrm{trigger}^\mathrm{SNR}=100$. 
The latter case is intended to account for surrounding noise from the horizon, as discussed in \cite{Jannis}.
The vertical black dashed line in the left part of Fig.\,\ref{fig:RadioSignal} represents the maximum of the absolute the electric field, indicating the timing $t_i$ when the radio signal from the position of the shower maximum $X^{\text{radio}}_{\text{max}}$ arrives at the antenna.

%If the fired number $N_\mathrm{fire}\ge4$, the radio array is triggered. 
An event accepted by the data acquisition (DAQ) is assumed to require at least three triggered antenna stations, i.e.\ $N_\mathrm{trigger} \ge 3$.
Figure~\ref{fig:RadioSignal} (rhs) shows the radio trigger efficiencies compared with the water-Cherenkov detector (WCD) at the Pierre Auger Observatory \cite{icrc07} for 1\,EeV and 10\,EeV with $\theta = 85^\circ$.
The abscissa is the $E_{\nu_e}^\mathrm{CC}$ interaction injection position $X_1$ in units of the slant depth, which denotes the integrated air density, written approximately as:
\begin{equation}
X_{1} \approx \int_{h_{\text{top}}}^{h_{1}} \frac{\rho(h)}{\cos\theta} \, dh \, ,
\end{equation}
where $h$ is the altitude above sea level and $\rho$ the density of the air. 
% $X_1=0$ represents the top of the atmosphere. The slant depth at the Auger site is shown in the figure.
%A curved atmosphere is considered in the right of figure~\ref{fig:RadioSignal}, as well as in the following work.
All simulations consider a curved atmosphere.

At $E_{\nu_e}^\mathrm{CC} = 1$\,EeV, the trigger efficiency is highly sensitive to the value of  T$_\mathrm{trigger}^\mathrm{SNR}$. For the conservative case (T$_\mathrm{trigger}^\mathrm{SNR}$=100), the  exposure of triggered events shrinks to less than that of the WCD.
For the higher energy case, $E_{\nu_e}^\mathrm{CC}=10$\,EeV, the exposure is always greater than that of the WCD, regardless of the trigger threshold. 
However, it is difficult to trigger the radio array for very young showers close to the ground because the showers need to be sufficiently developed to generate a detectable radio signal. Additionally, the cone-like development of the footprint must to be large enough to trigger a sufficient number of antennas spaced 1.5\,km apart.
Therefore, as shown on the right side of Fig.\,\ref{fig:RadioSignal}, the radio trigger efficiency is low compared with that of the particle detector.
Conversely, the radio signals in the 30-80\,MHz band can propagate much farther than particles can triggering the respective detectors. 
Therefore, we conclude that the complementary radio trigger exposure occurs at higher energies, 
$E \geq 10^{18}$\,eV, and with "older" showers than the particle trigger exposure.
Furthermore, the neutrino detection enhancement only occurs at inclined zenith angles where the radio footprint has enough space to to grow.
Therefore, the following study focuses on inclined angles beyond $75^\circ$. 

%air showers using CoREAS, applying:
%\begin{itemize}
%    \item A 30--80~MHz bandpass filter.
%    \item White noise consistent with measured RMS per polarization.
%    \item Antenna GPS time sync uncertainty of 5~ns.
%\end{itemize}

%A triggered antenna requires an electric field magnitude $|E|$ exceeding $3 \times 14~\mu$V/m, corresponding to $\text{SNR} > 10$. A valid event requires at least four triggered antennas.

\section{Neutrino detection scenario}

\begin{figure}[ht]
    \centering    \includegraphics[width=0.49\linewidth]{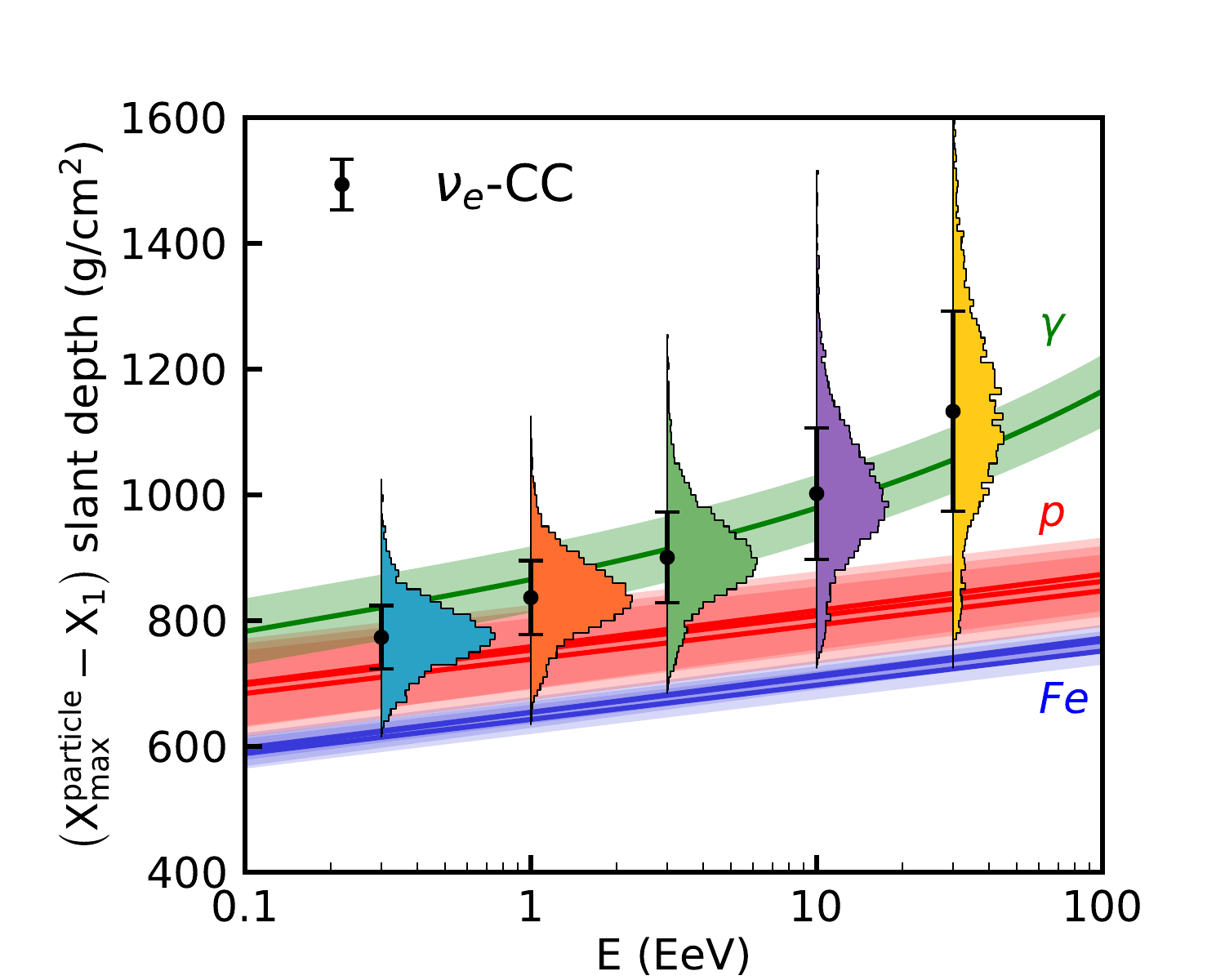}
    \includegraphics[width=0.49\linewidth]{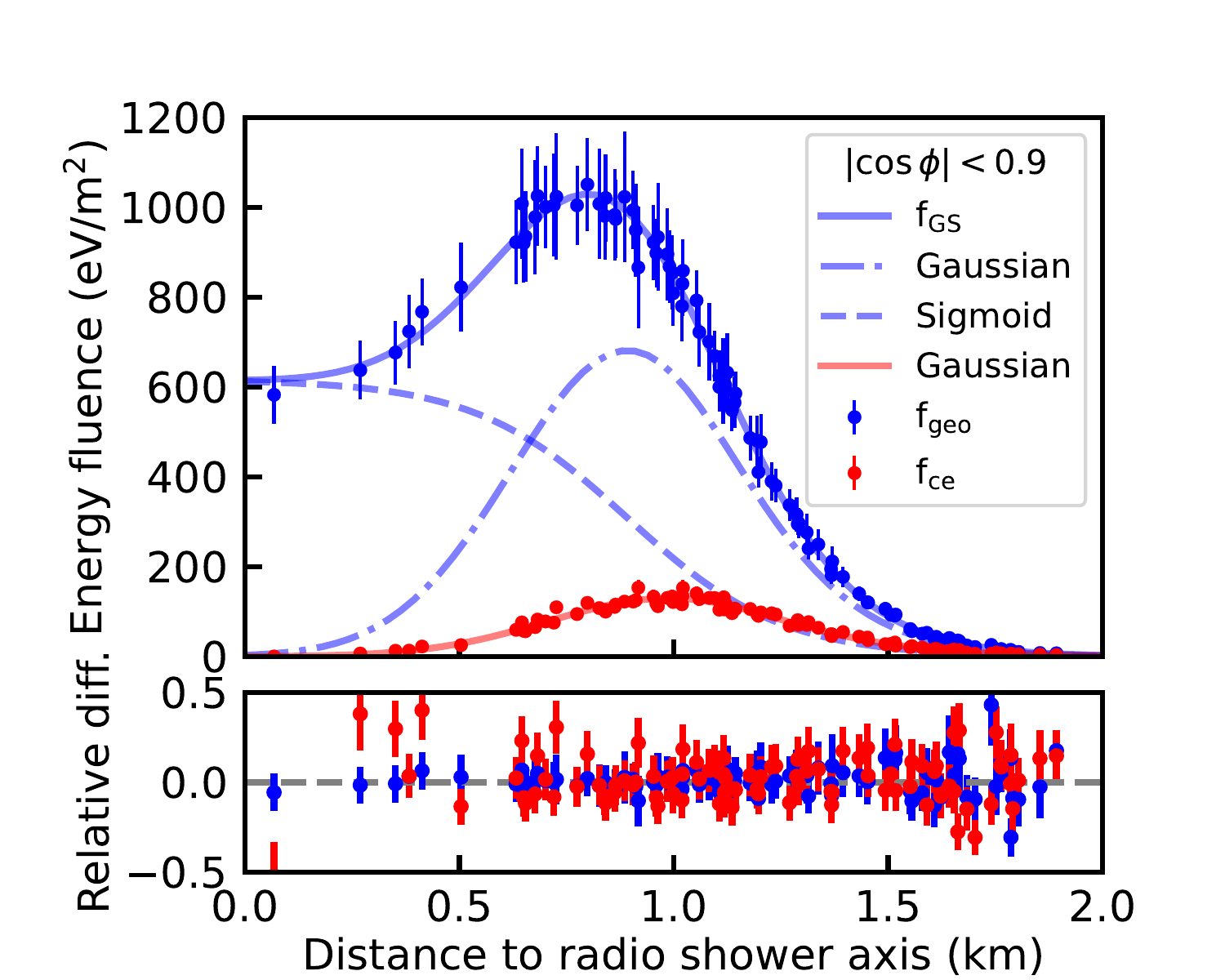}
    \caption{Left: the particle maximum ($X^{\text{particle}}_{\text{max}}-\text{X}_1$) distributions for proton, iron (EPOS-LHC, QGSJetII-04, and SIBYLL2.3d), photon ($\lambda=1$ in \cite{UHEphoton}), and $\nu_e$ (CC) as primary particles. 
    %The particle maximum of $\gamma$ follows the classical Gumbel distribution ($\lambda=1$) as described in \cite{UHEphoton}. 
    The distribution of $\nu_e$-CC is from very inclined simulations ($\theta>75^\circ$). Right: The performance of the footprint fits for a $\nu_e$-CC simulation of $E_{\nu_e}=30$ EeV, $\theta=87^\circ$, and $X_1=8000$\,g/cm$^{2}$ with 30-80\,MHz bandpass, white noise, and GPS timing uncertainty. $\phi$ represents the angle in the shower plane between the antenna position and the $\vec{v}\times\vec{B}$ axis ($\phi=0$: $\vec{v}\times\vec{B}$, $\phi=90^\circ$: $\vec{v}\times(\vec{v}\times\vec{B})$). }
    \label{fig:XmaxEnergyFluence}
\end{figure}

In this work, we propose using the shower depth of which the maximum of the radio emission is observed, $X_{\mathrm{max}}^{\mathrm{radio}}$, to distinguish neutrinos from UHECRs. 
This depth is defined as $X^{\text{radio}}_{\text{max}} = \int_{h_{\text{top}}}^{h_{\text{max}}} \frac{\rho(h)}{\cos\theta} \, dh$.
Due to the large interaction cross-section of the UHECRs (neutrons, charged CRs and photons), the depth of the first interaction, $X_1$, is expected to be less than 200\,g\,cm$^{-2}$ in the UHE domain.
%the position of first interaction $X_{1}$ should be at values smaller than $200$ g/cm$^2$ in the UHE domain.
%the $X_{\mathrm{max}}^{\mathrm{radio}}$ of these background will be close to the top of atmosphere.
Figure~\ref{fig:XmaxEnergyFluence} (left) presents the maximum of the particle density, $(X^{\text{particle}}_{\text{max}}-\text{X}_1)$ for $\nu_e$-induced showers (considering CC interactions only) in comparison to that expected for $\gamma$'s, proton and iron primaries.
For proton and iron primaries, $(X^{\text{particle}}_{\text{max}}-\text{X}_1)<1300$\,g\,cm$^{-2}$ is a conservative limit. 
Therefore, the absolute position $X^{\text{particle}}_{\text{max}}$ for $\nu_e$-induced showers will be required to be $X_{\text{threshold}} >1500$\,g\,cm$^{-2}$. 
Furthermore, the maximum radio emission position $X^{\text{radio}}_{\text{max}}$ is expected to be earlier than the particle maximum,  $X^{\text{particle}}_{\text{max}}$, as discussed in \cite{ChristianRadiation, FelixMarvin}. 
Therefore, $X^{\text{radio}}_{\text{max}}<X_{\text{threshold}}$ should be a stringent constraint for the charged CRs.
Since the upper limit of the flux of UHE $\gamma$'s is strongly constrained, we do not treat it as a neutrino background in this work.
For neutrinos, the first interaction $X_1$ can be anywhere.
Therefore, the absolute position $X^{\text{radio}}_{\text{max}}$ can be much larger than $X_{\text{threshold}}$, which can be used to identify neutrinos in the background caused by charged CRs.
%Once we found a event with the reconstructed $X^{\text{radio}}_{\text{max}}>X_{\text{threshold}}$ and the condition of $X^{\text{radio}}_{\text{max}} - X_{\text{threshold}} > 3\sigma_{X^{\text{radio}}_{\text{max}}}$, it should be neutrino.
The key neutrino identification criterion is the reconstructed $X^{\text{radio}}_{\text{max}}>X_{\text{threshold}}$ and the condition of $X^{\text{radio}}_{\text{max}} - X_{\text{threshold}} > 3\sigma_{X^{\text{radio}}_{\text{max}}}$.
Therefore, reconstructing the emission point $X^{\text{radio}}_{\text{max}}$ and the shower geometry, $\theta$, is essential to identifying neutrinos. 
A spherical fit is used to reconstruct the position of $X^{\text{radio}}_{\text{max}}$.
A 2D-fit of the radio energy fluence measured in the antennas provides the core of the footprint.
By linking the point of $X^{\text{radio}}_{\text{max}}$ and the core of the geomagnetic radiation footprint, we can determine the shower axis.

\subsection{Reconstruction of the maximum radio emission point $\vec{r}_{\rm max}$}
% $X^{\text{radio}}_{\text{max}}$
%Other studies \cite{LOPES} have pointed out that interferometric reconstruction have the potential to increase the reconstruction precision in the future \cite{washington+harm}.
%The key neutrino identification criterion is:\[X^{\text{radio}}_{\text{max}} - X_{\text{thresh}} > 3\sigma_{X^{\text{radio}}_{\text{max}}}\]
The position and emission time of $X^{\text{radio}}_{\text{max}}$ are represented by the vectors $\vec{r}_{\rm max}$ and $t_{\rm max}$, respectively.
The maximum radio pulse at the antenna corresponds to the maximum emission point, $X^{\text{radio}}_{\text{max}}$.
Assuming the maximum radio wavefront is spherical \cite{LofarSphericalFit}, we posit that it propagates to the antenna at a velocity of $\gamma c$ in the air. $\gamma$ is a factor that mimics the effects of the refractive index and the true wavefront geometry.
Therefore, the maximum $|E|$ timing $t_i$, for each triggered antenna, as shown as a vertical dash line in the left plot of Fig.\,\ref{fig:RadioSignal}, should be $t_\text{max}+\frac{|\vec{r}_\text{max}-\vec{r}_i|}{\gamma c}$. 
The $\chi^2$ function to estimate $\vec{r}_\text{max}$ can be written as 
\begin{equation}
    \chi^2=\sum_i \left( \frac{\left(t_i- t_\text{max}-\frac{|\vec{r}_\text{max}-\vec{r}_i|}{\gamma c}\right)}  {\sigma^\text{GPS}_t} \right)^2 \,.
    % + \left( \frac{1-\gamma}{\sigma_\gamma}\right)^2 \,.
\end{equation}
% where $\sigma_\gamma=\frac{c}{\Delta s}\cdot\sqrt{2}\cdot\sigma_t^\text{GPS}$ and $\Delta s$ is the distance between two antennas. 
In order to have a good estimation of the emission point $\vec{r}_\text{max}$, a quality cut of $\chi^2/ndf<10$ is applied.

\subsection{Reconstruction of the radio footprint}

The geomagnetic energy fluence is  radially symmetric in the shower plane ($\vec{v} \times \vec{B}$,  $\vec{v} \times (\vec{v} \times \vec{B})$), enabling us to determine the position of the shower core position, $\vec{r}_\text{core}$, at the ground level. 
Using the method described in \cite{jcap2023}, we can reconstruct the shower core by using the geomagnetic energy fluence via
\begin{equation}
\label{eq:fgeo}
    f_\text{geo} = \left(\sqrt{f_{\vec{v}\times\vec{B}}} - \frac{\cos\phi}{|\sin\phi|}\cdot \sqrt{f_{\vec{v}\times(\vec{v}\times\vec{B})}} \right)^2\,,  \quad  
   % f_{ce} = \frac{1}{\sin^2\phi}\cdot f_{\vec{v}\times(\vec{v}\times\vec{B})}
\end{equation}
where $\vec{B}=0.24$ G is the magnetic field with an inclination of approximately $-36^\circ$ at the Auger site, and $\vec{v}$ is the direction of the shower axis. 
The reconstructed position, $\vec{r}_\text{max}$, of $X^{\text{radio}}_{\text{max}}$ is an input for the shower axis $\vec{v}=\vec{r}_\text{core}-\vec{r}_\text{max}$. 
After correcting for the so called "early-late" effect and subtracting the charge excess component, the geomagnetic energy fluence can be expressed as a combination of Gaussian and a sigmoid functions:
\begin{equation}
    f_\text{GS} = f_0 \Bigg( \underbrace{  \exp\left( - \left(\frac{r-r_0}{\sigma}\right)^{p(r)}\right) }_\text{Gaussian} + \underbrace{\frac{a_\text{rel}}{1+\exp(s\cdot\left(r/r_0 - r_{02}\right))} }_\text{Sigmoid} \Bigg) \,,
\end{equation}
where $f_0$, $r_0$, $\sigma$, $p(r)$, $a_\text{rel}$, $s$, and $r_{02}$ are free parameters. In addition, $p(r)=2$ when $r<r_0$ and $p(r)=( (r_0/r)^{b/1000}$ when $r>r_0$.
%\left\{\begin{matrix} 2\quad \quad\quad\quad\quad\quad\,\,\,(r<r_0)\\2\cdot (r_0/r)^{b/1000} \quad (r>r_0)\end{matrix}$.
The charge excess energy fluence can be obtained via $ f_\text{ce} = \frac{1}{\sin^2\phi}\cdot f_{\vec{v}\times(\vec{v}\times\vec{B})}$. 
We found that the Gaussian parameterization part of Eq.~\ref{eq:fgeo} can also describe the charge excess energy fluence.
Figure~\ref{fig:XmaxEnergyFluence} (rhs) shows the performance of the footprint fits to a simulated event for both the geomagnetic and charge excess components. 

Since the geomagnetic radiation is the dominant component, we only use the geomagnetic radiation fit to estimate $\vec{r}_\text{core}$. 
A quality cut of $\chi^2/ndf<10$ is used to select good fits. 

\subsection{Shower axis reconstruction}
\begin{figure}[ht]
    \centering
    \includegraphics[width=0.455\linewidth]{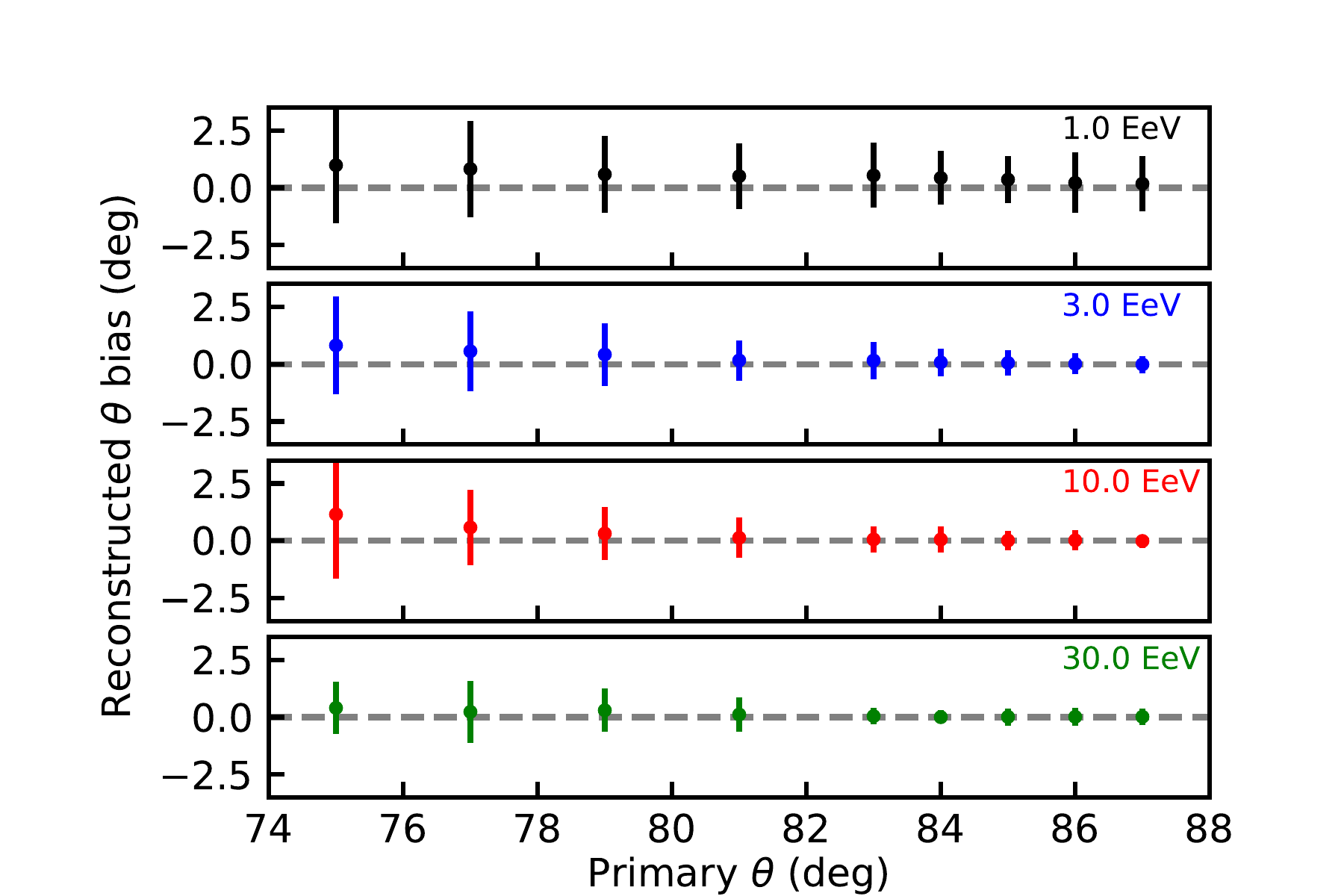}
    \includegraphics[width=0.49\linewidth]{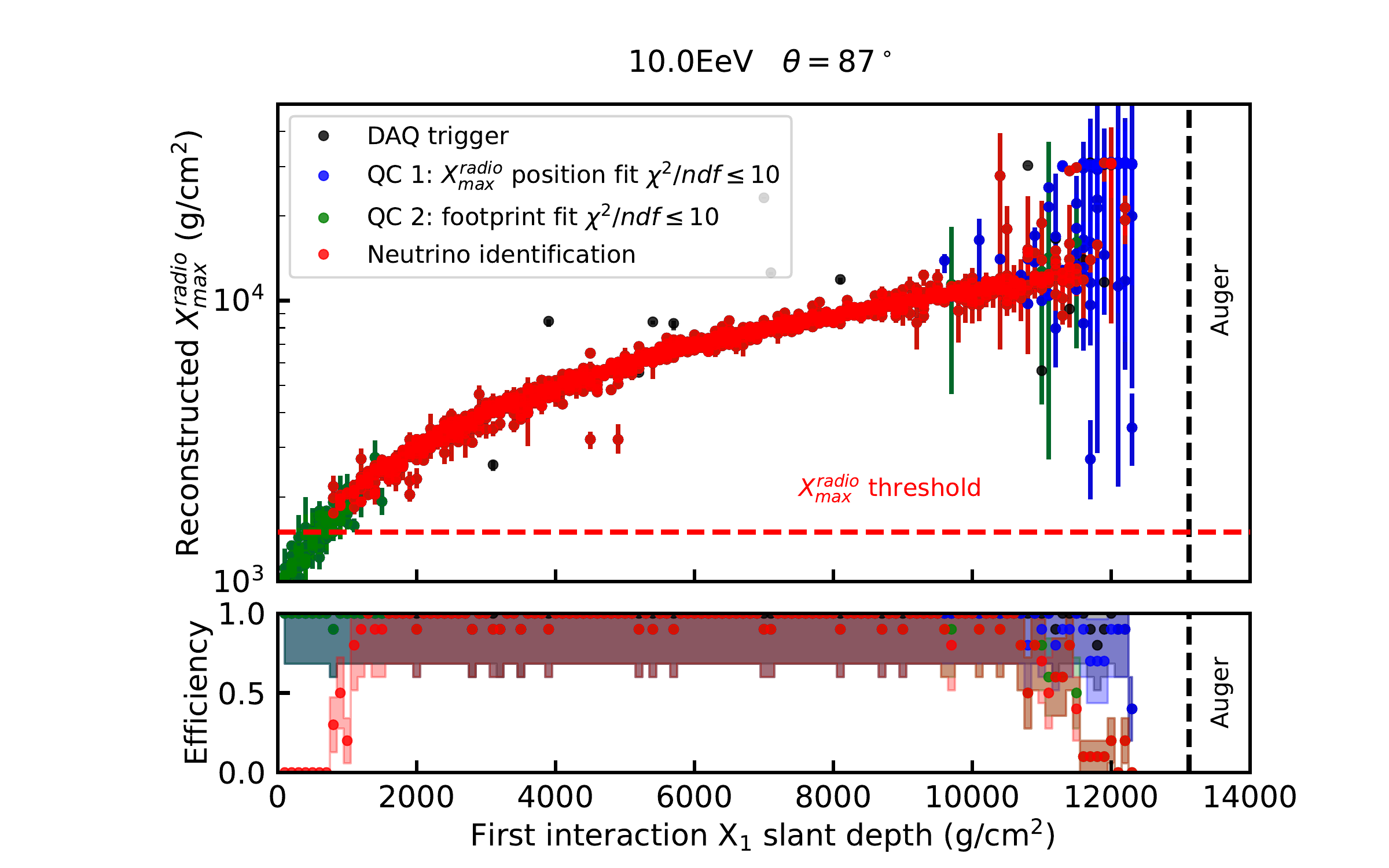}
    \caption{Left: the reconstructed zenith angle $\theta$ of the shower axis as a function of primary $\theta$ at given primary energy. Right: neutrino identification efficiency at $E_{\nu_e} = 10$\,EeV and $\theta=87^\circ$ as a function of the slant depth $X_1$ of the $\nu_e$-CC interaction for the case of T$_\mathrm{trigger}=50$.}
    \label{fig:Recon}
\end{figure}

To properly estimate the uncertainty of the shower axis, the consider the error propagation from the reconstructed $\vec{r}_\text{max}$ and $\vec{r}_\text{core}$.
Using two quality cuts (the spherical and footprint fits), we demonstrate the performance on the left of Fig.\,\ref{fig:Recon}.
%The left of figure~\ref{fig:Recon} shows the reconstructed $\theta$ as functions of primary $\theta$ and $E$.
%The error propagation from the reconstructed $\vec{r}_\text{max}$ and $\vec{r}_\text{core}$ to the reconstructed shower axis has been considered.
At a given primary zenith angle, $\theta$, the higher the energy, the lower the uncertainty of the reconstructed $\theta$.
At a given energy, the greater the zenith angle $\theta$, the lower is the bias and uncertainty of the reconstructed $\theta$.
Once the positions of $X^{\text{radio}}_{\text{max}}$ and the shower axis are known, the air density can be integrated from the top of the atmosphere to the maximum emission point, denoted by  $\vec{r}_\text{max}$. We have considered the error propagation from the shower axis ($\theta$) and the maximum emission point $\vec{r}_\text{max}$ to $X^{\text{radio}}_{\text{max}}$.

%Therefore, the position of the maximum emission point $\vec{r}_\text{max}$ and the shower axis ($\theta$) are needed. 
%We reconstruct the shower axis via spherical and footprint fits, applying quality cuts of $\chi^2/\text{ndf} < 4$ (spherical) and $< 10$ (footprint).

%The key neutrino identification criterion is:\[X^{\text{radio}}_{\text{max}} - X_{\text{thresh}} > 3\sigma_{X^{\text{radio}}_{\text{max}}}\]

%Where $X_{\text{thresh}}$ is derived from simulations of hadronic showers.

%\section{Energy Reconstruction and Fluence Modeling}

%We model the radio radiation energy as a function of shower energy using a geo-signal fluence model. The fit accounts for antenna positions, geomagnetic orientation ($\vec{v} \times \vec{B}$), and the relative angle $\phi$.

%Events near the $\vec{v} \times (\vec{v} \times \vec{B})$ axis present low signal strength and require special handling.

\section{$\nu_e$-CC detection efficiency and the effective area}

Figure~\ref{fig:Recon} (rhs) illustrates the neutrino identification efficiency as a function of the slant depth $X_1$ of the $\nu_e$-CC interaction after applying the trigger condition, quality cuts (QC), and neutrino identification cuts, i.e.\ $X^{\text{radio}}_{\text{max}} - X_{\text{threshold}} > 3\sigma_{X^{\text{radio}}_{\text{max}}}$, respectively.
The identification efficiency is very high when the shower is fully developed at such energy. 
However, when the showers are young and close to the ground, the spherical wavefront and the radio footprint model are ineffective. 
In these cases, particle detectors will be more effective at searching for neutrinos unless we improve on the description of young showers with radio antennas.

\begin{figure}[ht]
    \centering
    \includegraphics[width=0.45\linewidth]{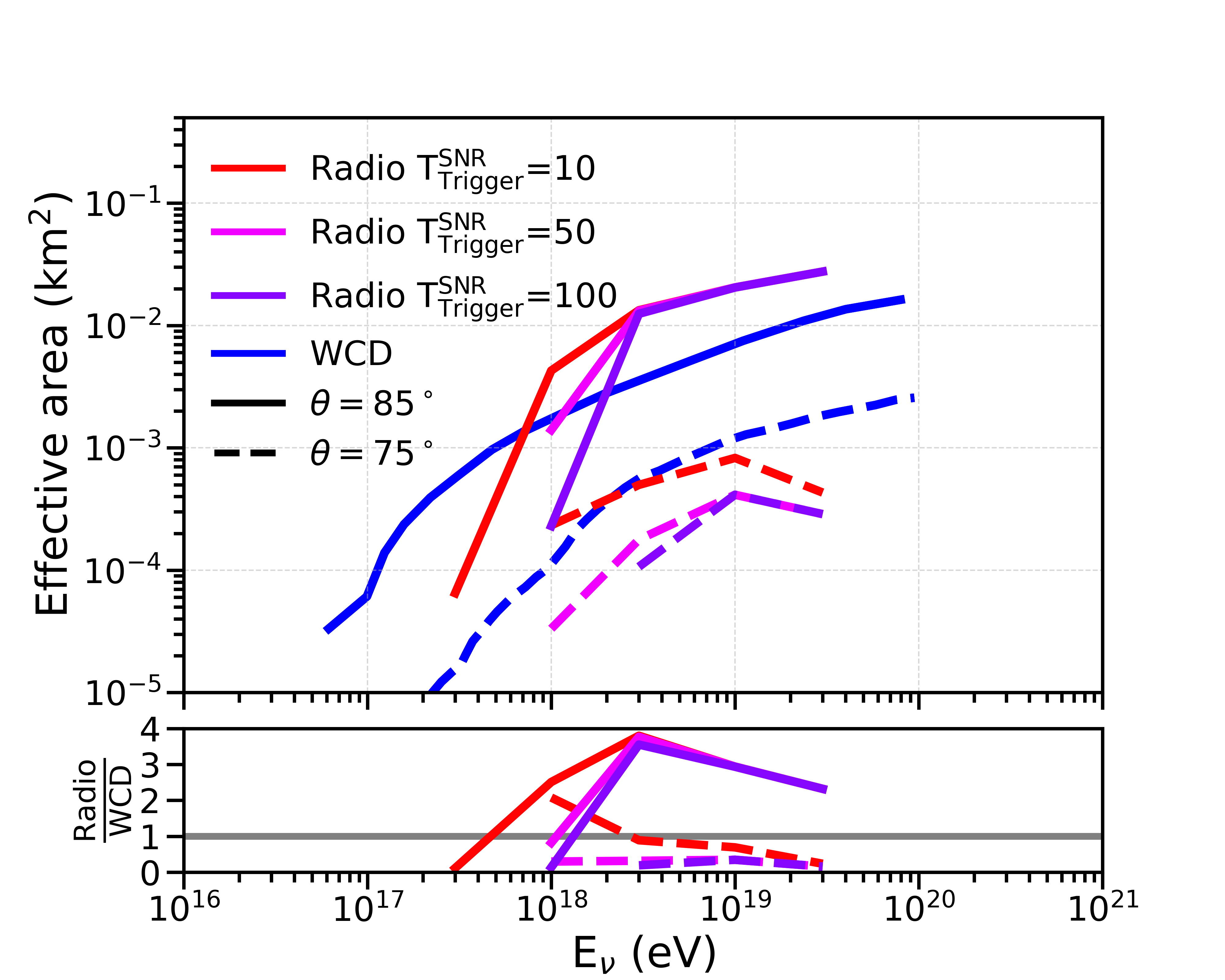}
    \includegraphics[width=0.45\linewidth]{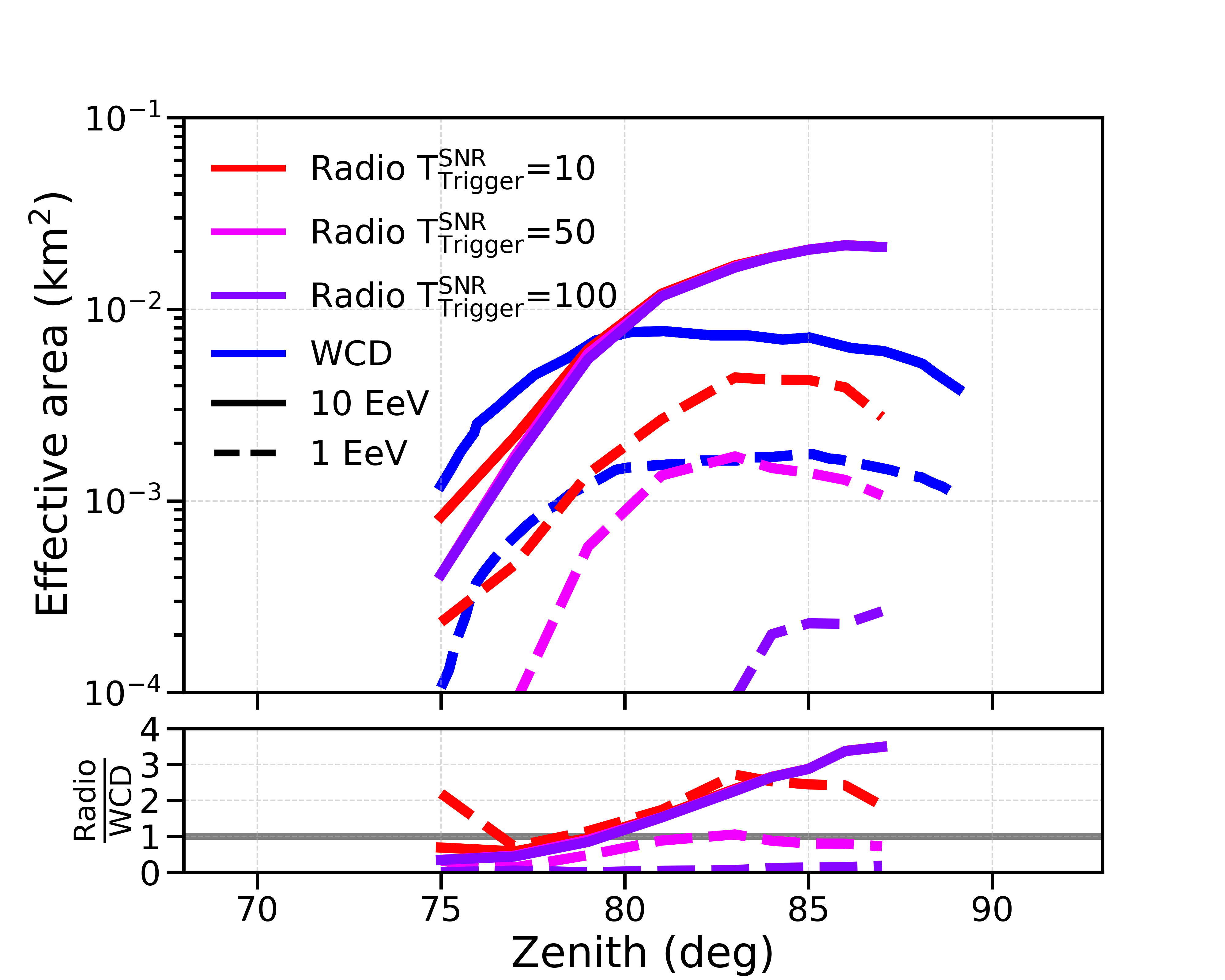}
    \caption{Left: the $\nu_e$-CC effective area as a function of primary energy. Right: the $\nu_e$-CC effective area as a function of the angle of inclination.}
    \label{fig:EffectiveArea}
\end{figure}

The overall detection efficiency, $\epsilon(X_1)$, includes trigger efficiency, reconstruction quality, and neutrino identification selection.
The effective area of the $\nu_e$-CC detection is computed as follows:
\begin{equation}
    A_{\text{eff}} = \iint \cos\theta \, \epsilon(X_1) \, \sigma_{\nu_e}^\text{CC} \, m_p^{-1} \, dX_1 \, dA \,,
\end{equation}
where $\sigma_{\nu_e}^\text{CC}$ is the cross-section of $\nu_e$-CC, $m_p$ is the proton mass, and $A$ is the area of the radio antennas ($A=3000$\,km$^2$ for Auger).
Figure~\ref{fig:EffectiveArea} shows a comparison of the effective area of a radio antenna array to that of WCDs \cite{jcap2019}. 
The plots show that the radio detection becomes dominate when the energy is greater than 1 EeV for showers with an inclination angle of $\theta>80^\circ$.
%The right plot shows that the effective area of radio detection 
At high inclinations and energies, radio detection provides larger effective area, up to a four times larger than that of WCDs.

%\section{Discussion and Outlook}

%The study confirms that radio detection enhances the sensitivity to inclined neutrino-induced showers. The radio-based reconstruction of $X^{\text{radio}}_{\text{max}}$ serves as a powerful discriminant against background hadronic showers. Future efforts will refine modeling for different flavors (NC interactions), improve the treatment of young showers, and extend the sensitivity to events outside the array.

\section{Summary and Outlook}

The detection of ultra-high-energy (UHE) neutrinos is one of the major frontiers in astroparticle physics. In this study, we have demonstrated the feasibility of using ground-based radio antennas to detect inclined neutrino-induced air showers. Specifically, we introduced an algorithm to reconstruct the maximum of the radio emission ($X^{\text{radio}}_{\text{max}}$), enabling us us to efficiently distinguish between EAS induced by neutrinos and cosmic-rays. 
% This method improves the ability of ground-based observatories to detect neutrinos, especially the very inclined EAS above 1~EeV.
% where traditional particle detectors are limited by atmospheric attenuation.
 % and the geometry of the shower axis

Using the Pierre Auger Observatory as a reference detector, we have demonstrated that radio arrays provide improved effective area for the down-going $\nu_e$-CC induced showers at energies above 1~EeV and zenith angles beyond $75^\circ$, complementing the existing surface detectors.
% Our simulation studies confirmed that radio detection becomes increasingly effective at energies above 1~EeV and zenith angles beyond $75^\circ$. 
Furthermore, the $X^{\text{radio}}_{\text{max}}$ observable is a powerful tool for rejecting cosmic-ray backgrounds. 
% based on their typically shallow development profiles.

% These techniques are not limited to Auger. 
These techniques can also be used at other ground-based radio observatories, such as GRAND ~\cite{GRAND}.
% The next generation of radio observatories, such as the Giant Radio Array for Neutrino Detection (GRAND)~\cite{GRAND}, aims to deploy tens of thousands of antennas across tens of thousands of square kilometers. 
The reconstruction and identification techniques developed in this work can be applied directly or adapted for such large-scale instruments. 
% Similar opportunities exist for other experiments that employ radio detection of inclined air showers, including BEACON~\cite{BEACON}, and TAROGE~\cite{TAROGE}.
%, and RNO-G~\cite{RNOG}. 

Future work will expand upon this method by including neutral-current (NC) and other flavor interactions, developing more robust approaches for handling young showers close to the ground, and testing the sensitivity of sparse and hybrid radio arrays. 
The influences from the radio antenna response, timing synchronization will also be tested.
% In parallel, improvements in radio hardware, timing synchronization, and noise suppression will expand the capabilities of this technique. 
% These efforts will pave the way for the discovery of the first EeV neutrinos and the identification of their astrophysical sources.
The radio technique will pave the way for the detections of the EeV neutrinos and the identification of their astrophysical sources.

\section*{Acknowledgements}
We would like to thank our colleagues in the radio group within the Pierre Auger Collaboration for the valuable suggestions and fruitful discussions. 
We are grateful to the Pierre Auger collaboration for simulation resources with $\nu_e$-CC. 
The calculation for this work is performed on the pleiades Cluster at University of Wuppertal.
This work was supported by the University of Wuppertal, the Deutsche Forschungsgemeinschaft (DFG), and the BMBF Verbundforschung Astroteilchenphysik.

%\section{Neutrino indued air showers}

%\section{Radio signal and antenna response}

%\section{Reconstruction of neutrino induced shower with radio antenna}

%\section{Identification of neutrinos}

%\section{Effective area}

%\section{Surmmary}

\end{document}